\documentclass[english,aps,twocolumn]{revtex4}
\usepackage[T1]{fontenc}
\usepackage[latin9]{inputenc}
\usepackage{graphicx}

\makeatletter

\usepackage{babel}
\makeatother

\begin{document}

\title{A Minimal Model for the Study of Polychronous Groups}

\date{March 27, 2008}

\author{Willard L. Maier}
\email{willardmaier@sbcglobal.net}
\affiliation{Texas Christian University }

\author{Bruce N. Miller}
\email{b.miller@tcu.edu}
\affiliation{Texas Christian University}

\begin{abstract}
A minimal model of polychronous groups in neural networks is
presented. The model is computationally efficient and allows the
study of polychronous groups independent of specific neuron models.
Computational experiments were performed with the model in one- and
two-dimensional neural architectures to determine the dependence of
the number of polychronous groups on various connectivity options.
The possibility of using polychronous groups as computational
elements is also discussed.
\end{abstract}
\maketitle

\section{Introduction}

Significant progress has been made in understanding the human brain
over the past half century. The behavior of individual neurons has
been studied extensively, using both experimental and computational
methods, to the point where science can explain not only the
characteristics of the various neuron types within neural networks,
but can also give a detailed account of the mechanisms within the
neurons themselves that cause these behaviors. Despite this
progress, there is still a huge gap in our understanding of how
these low-level mechanisms eventually result in the high-level
cognitive functions of the brain.

One phenomenon whose understanding may help bridge this gap is
\emph{polychronization}, an idea that was introduced by Izhikevich
in 2006 \cite{Izh06}. In a network with interconnection delays, two
neurons may fire at distinct times, yet have their spikes arrive at
a common postsynaptic neuron simultaneously due to the difference in
connection delays. This phenomenon is termed polychronization. In
addition these neurons plus the stimulated postsynaptic neuron may
have their output spikes arrive simultaneously at still other
neurons, causing further neural activity. The set of neurons in this
chain reaction is called a polychronous group, which we sometimes
shorten to \emph{polygroup}.

Polychronization is similar to the phenomenon of \emph{synfire
chains} \cite{Abeles91} \cite{Bienenstock95}. However synfire chains
appear when the neural network has synaptic connections with
identical delay times, whereas polychronization occurs when there is
a spectrum of connection delays between neurons, and is more like
the idea of a \emph{synfire braid} mentioned by Bienenstock. It has
been suggested that synfire chains form the basis of learning in the
neocortex \cite{Doursat06}, while others have explored the
information processing aspects of such chains \cite{Claussen06}. The
focus of this paper is on neural networks with transmission delays
between neurons, a necessary condition for the appearance of
polychronization.

Precisely timed spatiotemporal patterns have been observed
experimentally both \emph{in vivo} and \emph{in vitro}
\cite{Rolston07} \cite{Abeles93}. Although these experiments seem to
provide evidence for the existence of polychronous groups in the
brain, it is an open question as to whether such observed activity
can be accounted for by surrogate data generation. While detection
of polychronous groups in theoretical models is straightforward, the
lack of full network data in experimental situations makes their
observation problematic.

Izhikevich noted that the number of polychronous groups far exceeded
the number of neurons in the systems he studied. This observation
led him to hypothesize that polychronous groups may represent
memories in the brain, which could possibly explain the rich
diversity of brain behavior that seemingly transcends the
capabilities of the neurons present.

In this paper we describe a simple neural network model that has a
minimal number of features to support the study of polychronous
groups. We also develop an associated algorithm for the calculation
of polygroups formed in the model, and apply that algorithm to
various random networks to determine the number of potential
polygroups in these systems.

Additionally we describe a new form of neural computation using
polychronous groups as the basic computational elements. The
simultaneous firing of two polygroups can in some cases stimulate
the formation of still other polychronous groups, leading to a
cascade of activity extending far beyond the space and time of the
initial neural firings. This combination of polygroups into new
polygroups suggests a higher level structure to the dynamics of
neural systems.

\section{Description of the model}

\subsection{Network model}

In the original paper on polychronous groups, Izhikevich analyzed a
network of neurons modeled individually by his own spiking neuron
equations \cite{Izh03}; in addition, Spike Timing Dependent
Plasticity (STDP) was used to adjust the weights in the network.
Other researchers have also stressed the importance of STDP in
forming such groups \cite{Hosaka08} \cite{Izh04}. While these
features create a system that has certain characteristics of actual
neurons in the brain, they are not necessary to study the phenomenon
of polychronization. One of the key premises of this paper has been
to abstract the system to the bare minimum features necessary for
studying the pure computational concepts of polychronization and
polychronous groups.

A simple digraph with connection delays is sufficient to model the
essential features of polychronization. A neuron model that fires a
spike when the sum of its inputs reaches a fixed threshold is used
for the nodes of the digraph. Connections between neurons are
lossless, and each has a fixed, integer delay associated with it.
Discrete time is used in the model with the same integer scale. All
connections are excitatory in the basic model.

The model assumes that if a neuron receives two or more simultaneous
input spikes it will activate and fire its own spike. A system in
which a single input spike causes a neuron to fire cannot be
particularly interesting, since all that has happened in
computational terms is that the spike has been delayed. Requiring a
large number of simultaneous spikes for activation is more realistic
in terms of modeling the human brain; it has been estimated that it
takes 20 to 50 presynaptic spikes arriving within a short time
window to cause a postsynaptic spike in the human brain
\cite{Gerstner02}. However such a system would be far more difficult
to analyze, and is simply not necessary for understanding the
fundamentals of polychronization. Hence, requiring two spike
arrivals is the simplest and most tractable arrangement that will
yield computationally rich behavior.

To build a network in which to search for polychronous groups, we
first choose $N$, the number of neurons in the network, and arrange
these $N$ neurons in a circular array (i.e. a linear array with
periodic boundary conditions). To choose the interconnections
between these neurons two parameters are used, 1) a fixed number of
input connections per neuron $m$, and 2) a radius $r$ of nearest
neighbors of each neuron from which connections may be selected.
When selecting input connections the neuron itself is excluded since
we do not want self-connection. In our initial models, once the
connections are set, each is assigned an integer delay chosen
randomly from the range $[d_{min},d_{max}]$, where $d_{min}$ and
$d_{max}$ are parameters of the model.

Notice that once the neural topology is fixed, the set of
polychronous groups within the network is also fixed. The network
itself can be studied to determine what polygroups are inherent
within it, irrespective of any specific dynamic considerations.

\begin{figure}
\scalebox{0.45}{\includegraphics{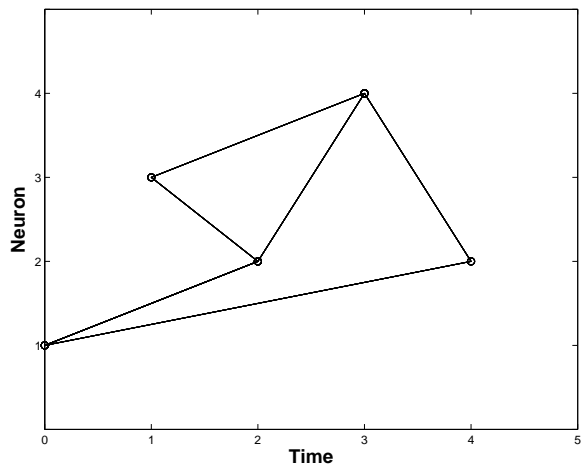}}
\caption{\label{fig:group1} A depiction of an example of a
polychronous group.}
\end{figure}

An example of a polychronous group is depicted in figure
\ref{fig:group1}. The vertical axis labels neurons and the
horizontal axis shows time. The circles mark points at which
specific neurons fire, and the lines show the travel of spikes from
left to right from one neuron to another. In this example the two
initiating neurons are neuron 1 which fires at t=0, and neuron 3
which fires at t=1. Spikes from these two neurons arrive at neuron 2
at time t=2, causing it to fire (this implies that the delay from
neuron 1 to neuron 2 is 2 time units, and the delay from neuron 3 to
neuron 2 is 1 time unit). Spikes from neuron 3 and neuron 2 arrive
simultaneously at neuron 4, causing it to fire at t=3. Finally,
spikes from neuron 1 and neuron 4 arrive at neuron 2 at t=4, causing
it to fire again.

\subsection{Finding Polychronous Groups}
A polychronous group is determined by the indices of its two
initiating neurons and the times at which they fire. The first step
in our search for polygroups is to scan through each possible pair
of neurons and examine each pair to see if it could initiate a
polygroup with an appropriate choice of firing times. For a system
with $N$ neurons we can form $N^2$ ordered pairs; however the
neurons must be distinct and their order is unimportant, so the
actual number of pairs we need to examine is $(N^2 - N) / 2$. For
each pair of neurons we must also choose the times at which they
fire. We are of course only interested in situations where these two
neurons will cause another neuron to fire; for this to happen they
must both have output connections to the same neuron. If such a
common postsynaptic neuron exists, it is always possible to choose
the initial firing times for the pair so that the postsynaptic
neuron receives spikes from them simultaneously. The times are
relative, allowing us to choose the earliest firing time to be $t=0$
and to choose the other time accordingly. For each pair of neurons
being considered, we must examine all possible connection pairs for
all common postsynaptic neurons.

The procedure above finds two neurons that stimulate a third neuron
to fire, but by definition to have a polygroup at least one other
neuron must also receive simultaneous spikes. The next step in the
algorithm is to search for additional firings by allowing the system
to evolve. The evolution of the system can be calculated efficiently
by creating a matrix of spike arrival counts (see expression
\ref{spikematrix} below). The rows are numbered from 0 to $N-1$
corresponding to the $N$ neurons in the system; the columns are
numbered from 0 to $t_{max}$, the maximum time to which the
simulation is run. The matrix entry $q_{n,t}$ represents the number
of spikes that arrive at neuron $n$ at time $t$. Initially we set
all matrix entries to zero, except for the two initial nodes which
we set to 2 at the appropriate times (this simulates these neurons
receiving 2 input spikes, so that they will fire during the
simulation run).

\begin{equation}
\label{spikematrix} \left[
  \begin{array}{ccccc}
    q_{0,0} & q_{0,1} & \ldots & q_{0,t_{max}-1} & q_{0,t_{max}} \\
    q_{1,0} & q_{1,1} & \ldots & q_{1,t_{max}-1} & q_{1,t_{max}} \\
    \ldots & \ldots & \ddots & \ldots & \ldots \\
    q_{N-2,0} & q_{N-2,0} & \ldots & q_{N-2,t_{max}-1} & q_{N-2,q_{0,t_{max}}} \\
    q_{N-1,0} & q_{N-1,0} & \ldots & q_{N-1,t_{max}-1} & q_{N-1,q_{0,t_{max}}} \\
  \end{array}
\right]
\end{equation}

To run the simulation we start at the leftmost column and look for
entries with a value of 2 or greater. These neurons have received
enough spikes to cause them to fire, so we look up what neurons they
are connected to along with the associated delays to find the times
at which the spikes arrive at the postsynaptic neurons. Using these
numbers we find and increment the corresponding matrix entries. We
then move to the second column and repeat the procedure.

The system can be run iteratively until no firings occur for a
period of time equal to the maximum delay in the system. However in
some cases polygroups can continue firing for a very long time; in
fact, polygroups can extend infinitely in time. For this reason a
limit is placed on how long the calculation will be performed. If
the limit is reached the group is flagged as being \emph{overrun} so
that subsequent analysis can take this into account. At the end of
the calculation, any matrix entry with a value of 2 or greater
corresponds to a firing neuron. If there are four or more such
entries, we have found a polygroup.

\section{Experimental Results}

\subsection{Results For One Dimensional Systems}

\begin{figure}
\scalebox{0.45}{\includegraphics{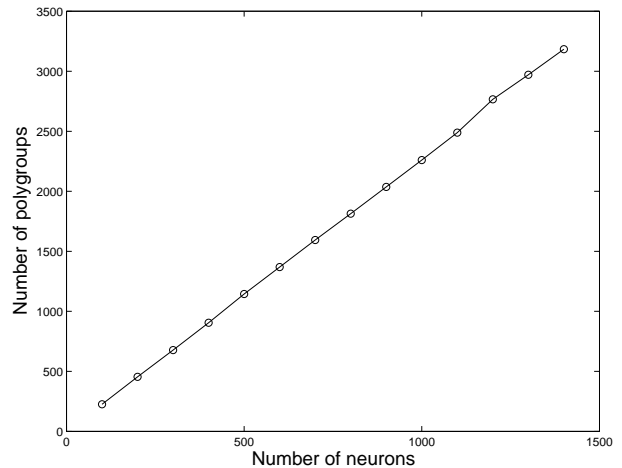}}

\caption{\label{fig:poly1} A plot showing the number of polychronous groups
as a function of $N$, the number of neurons in the system.}

\end{figure}

The parameters of the experiment that can be varied are:

\begin{enumerate}
\item $N$ = number of neurons in the network.
\item $m$ = number of input connections per neuron.
\item $r$ = radius of nearest neighbors of each neuron from which connections
may be chosen.
\item $d_{min}$ = minimum delay time.
\item $d_{max}$ = maximum delay time.
\end{enumerate}
The first set of experiments was set up to determine how the number
of polychronous groups varies as the the number of neurons in the
system is changed, holding all other parameters constant. Figure
\ref{fig:poly1} shows the results of these runs. For each value of
$N$, 30 runs were averaged together to give the mean number of
polychronous groups for that $N$. The relationship is clearly
linear, with a slope of about 2.2. For these runs, $m=5$, $r=5$,
$d_{min}=1$, and $d_{max}=5$.

\begin{figure}
\scalebox{0.45}{\includegraphics{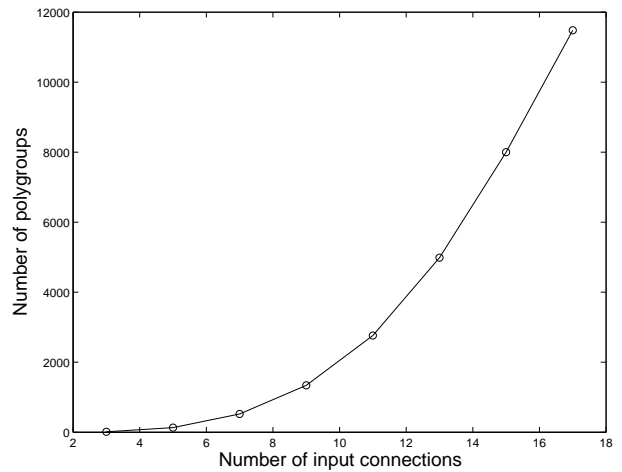}}

\caption{\label{fig:poly2} The number of polychronous groups as a function
of $m$, the number of input connections to each neuron in the network.}

\end{figure}

For the next experiments we decided to determine how the number of
polychronous groups varies as the the number of input connections to
each neuron changes. Results are displayed in Figure
\ref{fig:poly2}, which shows that the number of polychronous groups
increases rapidly as $m$ is increased. For these runs, $N=100$,
$r=10$, $d_{min}=1$, and $d_{max}=5$.

\begin{figure}
\scalebox{0.45}{\includegraphics{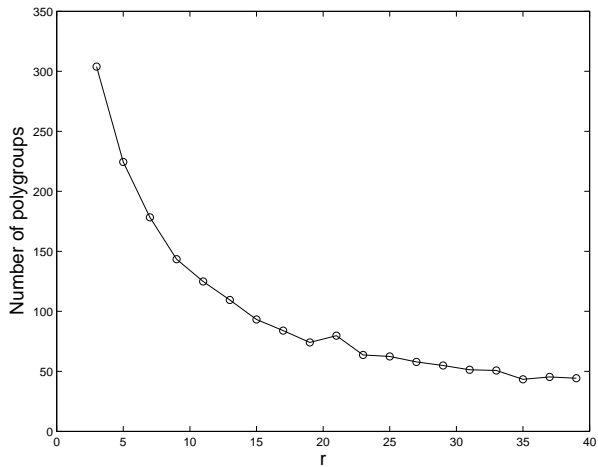}}

\caption{\label{fig:poly3} The number of polychronous groups as a function
of $r$, the radius of nearest neighbors from which input connections
are chosen.}

\end{figure}

We varied $r$ in the next set of experiments to determine how the
number of polychronous groups varies as the the number of nearest
neighbors from which input connections are chosen changes. Results
are displayed in Figure \ref{fig:poly3}, which shows that the number
of polychronous groups decreases rapidly as $r$ is increased. For
these runs, $N=100$, $m=5$, $d_{min}=1$, and $d_{max}=5$.

\begin{figure}
\scalebox{0.45}{\includegraphics{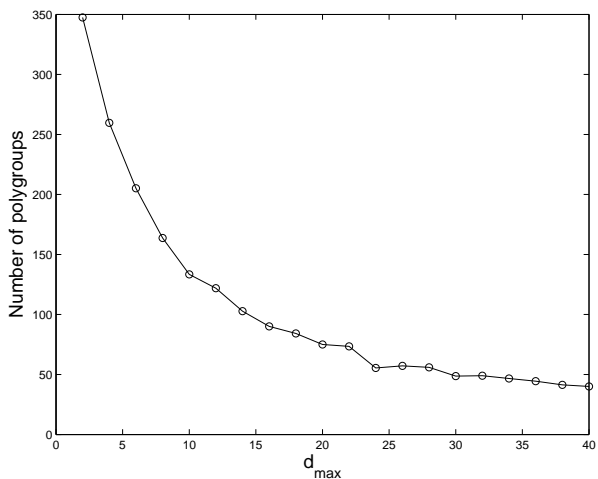}}

\caption{\label{fig:poly4} The number of polychronous groups as a function
of $d_{max}$, the maximum delay on a connection.}

\end{figure}

Figure \ref{fig:poly4} shows what happens as $d_{max}$ is varied.
The number of polychronous groups decreases rapidly as $d_{max}$
increases. This is intuitively clear when one considers that as $d_{max}$
increases, the number of possible delays on the connections increases
and so the probability of finding pairs of connections with simultaneous
arrivals becomes less. For these runs, $N=100$, $m=5$, $r=5$, and
$d_{min}=1$.

So what do these results tell us about polychronous groups in the
human brain? It is estimated that there are $10^{11}$ neurons in the
brain, with $m$ in the range of 1000 to 10,000 connections per
neuron. Connectivity in the neocortex has been observed to be about
10\%, so a good rough estimate of $r$ is 5,000 to 50,000.
Experimental measurements of axonal delays have shown that the delay
can be as low as 0.1 msec and as high as 40 msec \cite{Swadlow85}
\cite{Swadlow88} \cite{Swadlow92}. Since the number of polychronous
groups scales linearly with the number of neurons, we might expect
the number of groups to be roughly on the order of the number of
neurons. Of more concern, however, is the scaling relative to the
values of $r$, $m$, and $d_{max}$, since these scalings are
exponential in nature. Large values of $r$ and $d_{max}$ would tend
to lower the total number of polychronous groups, but a large value
of $m$ argues for a high number of such groups. The actual result
for the human brain cannot even be estimated with the numbers we
have so far.

Though the calculation of polychronous groups stretches the
capability of current computers, it is possible to define relatively
small networks and try to extrapolate measurements on them to
networks of a more realistic size. As a baseline we chose a system
with the parameters $N=5000$, $m=100$, $r=500$, $d_{min}=1$, and
$d_{max}=40$, which took about 12 hours of CPU time to run. For this
system there was a total of slightly more than $6.1\times10^{6}$
polychronous groups. If we were to estimate the number of
polychronous groups for this system based solely on the graph in
Figure \ref{fig:poly1}, we would expect somewhat over 10,000 groups;
the much larger actual total appears to indicate that the
exponential growth of the number of polychronous groups due to the
increase of $m$ overpowers the decrease brought about by the change
due to $r$ and $d_{max}$. This result agrees with that found by
Izhikevich in his original paper \cite{Izh06}.

\subsection{Results For Two Dimensional Systems}

Because of the brain's layered geometry, it is worthwhile to
investigate how dimensionality influences the availability of
polychronous groups. Here we address whether or not extending the
network to a two-dimensional topology affects the number of
polychronous groups. To answer this question both one and
two-dimensional networks were constructed using identical
parameters. For the two dimensional model, neurons were located on a
rectangular grid. The $m$ connections to a given neuron were
selected at random within a circle of radius $r$. The parameters for
the 1D and 2D networks were selected so that the same number of
neurons $N_{r}$ would be included in each sub-region of radius $r$.

\begin{figure}
\scalebox{0.45}{\includegraphics{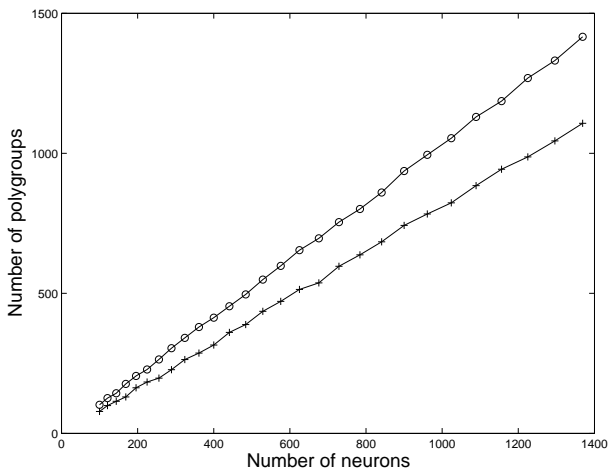}}

\caption{\label{fig:poly2d1} The number of polychronous groups as a function
of $N$, the number of neurons in the system, for networks with one-
and two-dimensional connectivity. The one-dimensional network graph
is marked by small circles, the two-dimensional graph by plus signs.}

\end{figure}

The variation of the number of polychronous groups as $N$ changed
is shown in Figure \ref{fig:poly2d1}. Both relationships are clearly
linear, though in the two-dimensional case the number of polychronous
groups is somewhat less. The slope of the 1D line is approximately
1.0, while the slope for the 2D line is about 0.8. Parameters for
these runs were $m=4$, $N_{r}=8$, $d_{min}=1$, and $d_{max}=5$.

\begin{figure}
\scalebox{0.45}{\includegraphics{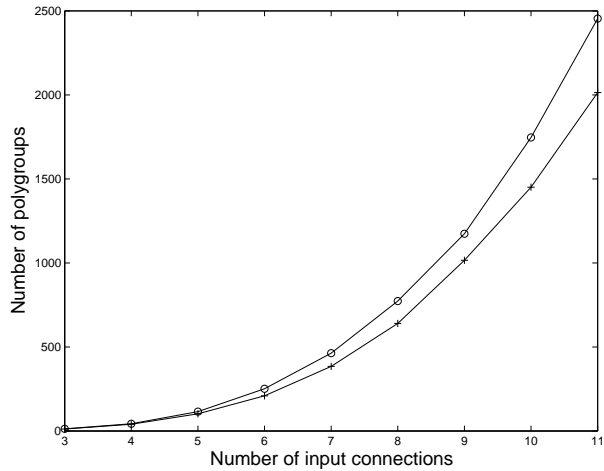}}

\caption{\label{fig:poly2d2} The number of polychronous groups as a function
of $m$, the number of input connections per neuron, for networks
with one- and two-dimensional connectivity. The one-dimensional network
graph is marked by small circles, the two-dimensional graph by plus
signs.}

\end{figure}

Figure \ref{fig:poly2d2} shows how the number of polychronous groups
depends on $m$, the number of input connections. Other parameters
were $N=100$, $N_{r}=24$, $d_{min}=1$, and $d_{max}=5$.

\begin{figure}
\scalebox{0.45}{\includegraphics{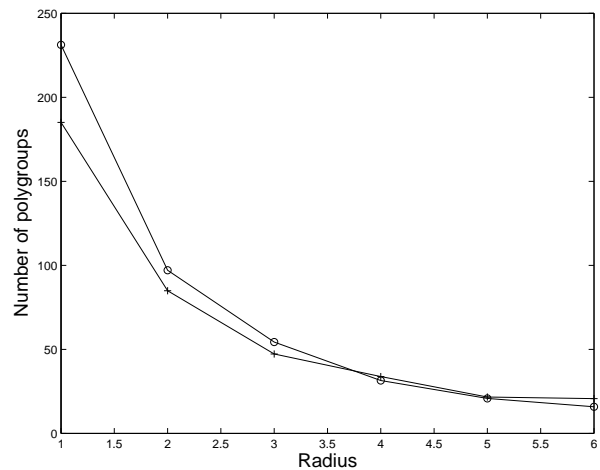}}

\caption{\label{fig:poly2d3} The number of polychronous groups as a function
of $r$, the radius from which input connections are selected, for
networks with one- and two-dimensional connectivity. The one-dimensional
network graph is marked by small circles, the two-dimensional graph
by plus signs.}

\end{figure}

Figure \ref{fig:poly2d3} shows how the number of polychronous groups
depends on $r$, the radius from which input connections are selected.
Other parameters were $N=225$, $m=4$, $d_{min}=1$, and $d_{max}=5$.

\begin{figure}
\scalebox{0.45}{\includegraphics{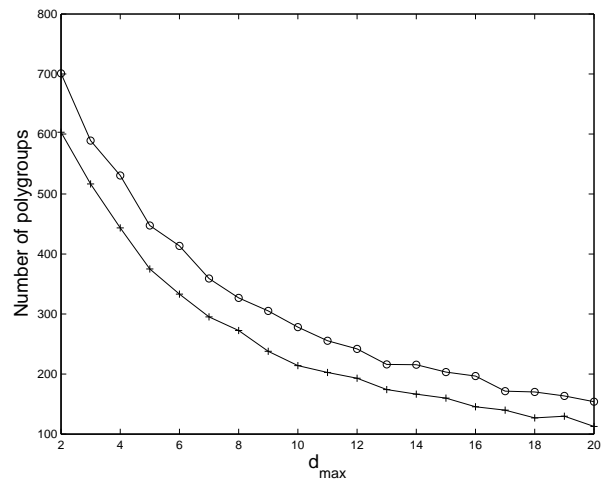}}

\caption{\label{fig:poly2d4} The number of polychronous groups as a function
of $d_{max}$, the maximum connection delay, for networks with one-
and two-dimensional connectivity. The one-dimensional network graph
is marked by small circles, the two-dimensional graph by plus signs.}

\end{figure}

Figure \ref{fig:poly2d4} shows how the number of polychronous groups
depends on $d_{max}$. Other parameters were $N=100$, $m=4$, $r=2$,
and $d_{min}=1$.

As can be seen from Figures \ref{fig:poly2d1} through \ref{fig:poly2d4},
the qualitative results for one and two dimensions are similar. The
actual number of polychronous groups does vary somewhat with each
of the parameters, but not significantly so. The net result of these
studies is that changing from one to two dimensions does not change
the essential form of the parametric dependencies.

\subsection{Choosing Connection Delays Deterministically}

In the simulations above the connection delays were chosen randomly
within a fixed range. It is a reasonable assumption, however, that
in actual networks of neurons the time delay associated with a synaptic
connection will be approximately proportional to the distance between
the connected neurons. If we use this assumption in our simulations,
how does it affect the number of polychronous groups in the network?

\begin{figure}
\scalebox{0.45}{\includegraphics{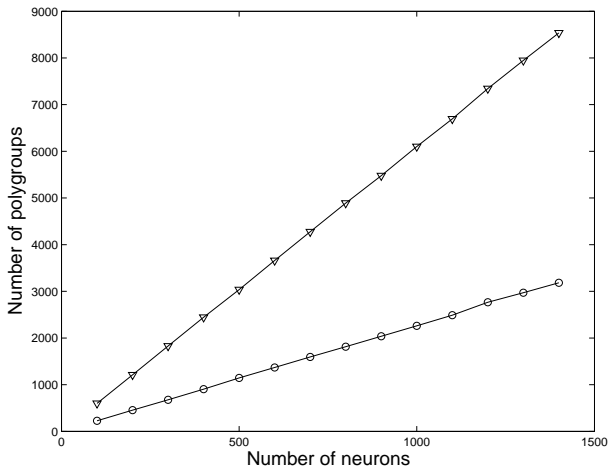}}

\caption{\label{fig:polyH1} The number of polychronous groups as a function
of $N$, the number of neurons in the system, for networks with random
vs. deterministic delays. The random network graph is marked by small
circles, the deterministic graph by inverted triangles.}

\end{figure}

\begin{figure}
\scalebox{0.45}{\includegraphics{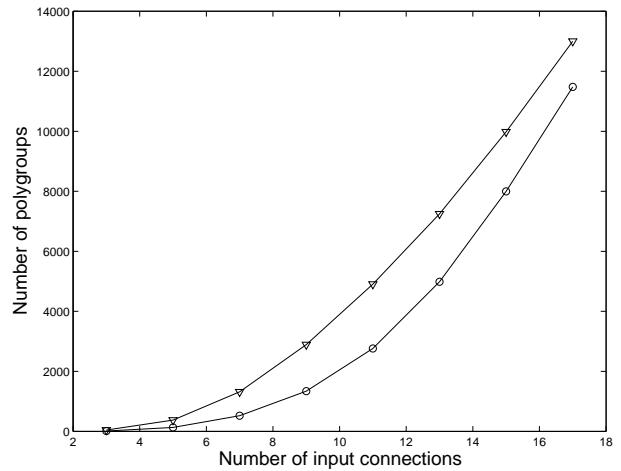}}

\caption{\label{fig:polyH2} The number of polychronous groups as a function
of $m$, the number of input connections per neuron, for networks
with random vs. deterministic delays. The random network graph is
marked by small circles, the deterministic graph by inverted triangles.}

\end{figure}

\begin{figure}
\scalebox{0.45}{\includegraphics{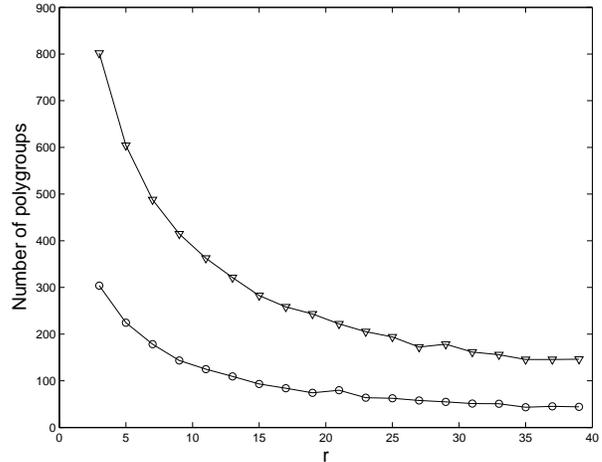}}

\caption{\label{fig:polyH3} The number of polychronous groups as a function
of $r$, the radius from which input connections are selected, for
networks with random vs. deterministic delays. The random network
graph is marked by small circles, the deterministic graph by inverted
triangles.}

\end{figure}

Figure \ref{fig:polyH1} shows how the number of polygroups varies
with $N$, for both a network with random delays and a network with
deterministic delays. Figures \ref{fig:polyH2} and \ref{fig:polyH3}
show how the number of polygroups varies with $m$ and $r$,
respectively. The essential form of the relationships do not change
when using deterministic delays, but the number of polygroups in the
networks with deterministic delays is significantly higher than the
corresponding networks with randomized delays. It is not immediately
clear why the number of groups increases when the delays are
proportional to the distance between the connected neurons.

\subsection{How Many Pairs Form Polygroups?}

Any given pair of neurons in our neural networks may or may not be
capable of stimulating a polychronous group, depending on their
synaptic connections and the associated delays, so it is reasonable
and interesting to calculate what fraction of the neuron pairs can
actually form polygroups. For a neuron with $m$ input connections
the number of possible ways that a pair of connections can be chosen
is given by
\begin{equation}
  \frac{m!}{2!(m-2)!}.
\end{equation}
Each pair of input connections has only one timing sequence with
which it will trigger the neuron, so this is also the number of ways
a particular neuron can be stimulated to fire. For a system with $N$
neurons, the total number of ways for neurons to be stimulated is
thus
\begin{equation}
  N\frac{m!}{2!(m-2)!}.
\end{equation}
Dividing the number of observed polygroups by this number gives us
the fraction of pairs that actually created a polygroup. Using the
data in Figure \ref{fig:polyH1} for the networks with deterministic
delays $(m=5,r=5)$, we find that the fraction of pairs that
stimulate polygroups is roughly constant over all $N$, and is equal
in this case to approximately 0.6. This may provide a lower bound
for more realistic systems where connections are correlated.

\section{Computation with polygroups}

A polychronous group can be thought of as a sort of automaton;
starting with just two firing neurons, an entire chain of neurons is
caused to fire over an extended period of time. The group is simply
a response to the initial stimulus, and in our perfect simulation
world of discrete time and distinct spikes, the response is
unvarying. In that sense, then, a polychronous group can be thought
of as a monolithic computational element.

When a polygroup is activated, the firing neurons within the group
will in most cases have connections to other neurons outside the
group. These outside neurons receive only a single spike and thus
will not fire. We can envision a "cloud" of such neurons surrounding
a polygroup in both space and time.

If two separate polygroups are activated whose firings overlap in
time, certain neurons in the surrounding clouds may receive two
simultaneous spikes, one from each polygroup, and thus be caused to
fire. Furthermore, two or more neurons may be activated in this
manner, and their combined action may in turn activate a totally
separate polygroup. The net result is that in some cases, the
activation of two polygroups can in turn activate a third polygroup.

For a given network, we can label each polygroup with an index $i$
and represent an arbitrary group with the symbol $G_i$. To fully
specify a polygroup we must know the time at which the group was
activated; since the relative times of the activating spikes are
fixed, we can choose the time of the first activating spike as the
time associated with the polygroup, and thus write $G_i^t$ to
indicate polygroup $i$ activated at time $t$.

If polygroup $G_1$ fires at time $t_1$ and polygroup $G_2$ fires at
time $t_2$, and if the combined action of these two groups causes
another polygroup $G_3$ to fire at time $t_3$, we can write
\begin{equation}
\label{polyadd1}
  G_1^{t_1} + G_2^{t_2} \rightarrow G_3^{t_3},
\end{equation}
where the symbol $\rightarrow$ is read "activates".

Times are all relative in the system, so any time offset $\tau$ may
be added without changing the above relationship:
\begin{equation}
\label{polyadd2}
  G_1^{t_1+\tau} + G_2^{t_2+\tau} \rightarrow G_3^{t_3+\tau},
\end{equation}

If indeed a polychronous group represents a memory in the brain,
then equation \ref{polyadd1} signifies that certain pairs of
memories are capable of stimulating a third memory. Equation
\ref{polyadd2} shows that the relation of these memories are time
invariant. It is interesting, however, that the two stimulating
memories must be activated in a fixed time relationship to each
other to cause the third memory to activate.

\section{Summary and conclusions}

We have developed a computationally efficient model for the study of
polychronous groups, constructed on the principle of including only
the essential features required for such groups. An algorithm is
included in the model to rapidly identify polygroups in the network.
The model was used to computationally investigate properties of
polychronous group formation in various network topologies.

Through numerical experiments we found that the number of polygroups
in the network depends linearly on the number of neurons, holding
all other criteria constant. The number of polygroups decreases
asymptotically as the radius of connectivity or the range of time
delays increases, but grows exponentially as the number of input
connections increases. By testing a larger system we found that the
exponential growth of the number of polygroups due to an increase of
input connections dominated over the other factors we studied.

We conducted similar experiments comparing one- and two-dimensional
networks, and found slight numerical but no qualitative differences
in the results. Experiments were then performed in which the
transmission delays were chosen to be proportional to the distance
between neurons, and when these results were compared with our
initial model we discovered that there were no qualitative
differences, but that the number of polygroups was much higher for
the network with the proportionally chosen delays.

We also introduced the concept of computation using polygroups. In
some cases two activated polygroups can cause the stimulation of a
third polygroup. This opens up the possibility of polygroups being
used as monolithic interacting elements in a neural system. Further
work is required to determine the properties of this type of
computation.

There are still many open questions regarding polychronous groups,
and we have only begun to explore their properties. Further
measurements could prove useful, such as determining the
distribution of the number of neurons per group under various
network topologies. Specific examples that have recently shown a lot
of interest are small world and scale free networks \cite{Albert02}.
Inhibition could also be added to the neural connections, to bring
the model more in line with the workings of biological neural
systems.

\bibliographystyle{plainnat}
\bibliography{Neuron}

\end{document}